\def\prd{Phys.~Rev.~D}
\def\be{\begin{equation}}
\def\ee{\end{equation}}      
\def\ba{\begin{eqnarray}}
\def\ea{\end{eqnarray}}
\def\mathrm{\rm}
\documentstyle[12pt]{article}
\def\rcorr{1}
\def\negm{2}
\def\negdiff{3}
\def\ppdiff{4}
\def\bediff{5}
\def\eic{6}
\begin{document}
\def\sls#1{/ \hskip -6pt #1}
\title{{The relevance of the vertex bremsstrahlung photon detection in the
$\nu_e(\overline{\nu}_{e})e^{-}$ 
scattering experiments at low energy}}

\author{
C.~Broggini \footnotemark[1]\, , I.~Masina
\thanks {Department of Physics Padova University and INFN sez.~ di Padova}\, ,
\\ 
M.~Moretti
\thanks {CERN - Theory Division, Geneva and 
Department of Physics Ferrara University and INFN sez.~ di Ferrara}
\\
}
\date{}
\maketitle
\begin{abstract}

We discuss the size of the 
$\nu_e(\overline{\nu}_{e})e^{-} 
\rightarrow \nu_e(\overline{\nu}_{e})e^{-}$
cross section reduction due to the rejection 
of the events with a vertex bremsstrahlung photon above a certain energy in 
the final state.
In particular we analyze 
the effect in experiments 
designed to detect the low energy 
$\overline{\nu}_{e}$ and $\nu_{e}$
from a nuclear reactor and from the Sun.
We find that such reduction has to be considered in a relatively
high statistic reactor experiment, while it is negligible for
$pp$ and $^{7}Be$ solar neutrino detection.

\end{abstract}

\section{Introduction}

Forthcoming reactor and solar neutrino experiments
are expected to provide  important improvements both in the quality
and in the statistics of the data. As a matter of fact,
in most cases the statistical error will be of a few per cent.
 At this level of accuracy, radiative 
corrections are likely to become important.

Our goal is to extimate the impact
for forthcoming experiments 
of the radiative corrections (see \cite{brc} for a review) to the process
$\nu_e(\overline{\nu}_{e})e^{-} 
\rightarrow \nu_e(\overline{\nu}_{e})e^{-}$
once
realistic experimental set-up are included into the analysis. In particular 
we will discuss the effect of the vertex bremsstrahlung photon in experiments 
designed to detect the low energy 
$\overline{\nu}_{e}$ and $\nu_{e}$
from a nuclear reactor and from the Sun.

\section {Radiative Corrections}

  A complete and detailed account of the full \cite{rc} set of 
one loop radiative corrections to $\nu_e e^-$ scattering,
as well as a comprehensive set of references, is given in
\cite{brc}.
The same corrections for $\bar \nu_e e^-$ scattering are obtained
by replacing $g_L(T)\leftrightarrow g_R(T)$ in the formulas given 
in \cite{brc}.
A preliminary study of the size of radiative corrections for a 
$\overline{\nu}_{e}e^{-}$ scattering experiment with the reactor 
antineutrinos, the MUNU one, has been done in \cite{noi} 
where we 
discussed the physics items which can be studied in a 
$\overline{\nu}_{e}e^{-}$ scattering experiment at a nuclear reactor in 
addition to the neutrino magnetic moment.

For the sake of completeness let us summarize a few items discussed in 
\cite{noi}:
\begin{itemize}

\item one loop radiative  corrections
induce a decrease, as compared to the Born prediction,
of about $2.5\%$, mildly dependent
on energy and scattering angle.  
In fig.~\rcorr\ we show the relative size of the order $\alpha$ 
cross section,
weighted over the antineutrino spectrum,  as a function of the 
electron scattering angle and kinetic energy;

\item higher order corrections are below the per mille level and
therefore, in view of the forecasted experimental accuracy, negligible.

\end{itemize}

\subsection {Bremsstrahlung corrections}

Radiative corrections as given in \cite{brc} are for
an {\em inclusive set-up}, {\it i.e.} they refer to the cross
section for the process 
$\nu_e(\overline{\nu}_{e})e^{-} 
\rightarrow \nu_e(\overline{\nu}_{e})e^{-}+(n\gamma)$
where with $+(n\gamma)$ we denote the sum over  
$\nu_e(\overline{\nu}_{e})e^{-}$
final
states with an arbitrary number of photons.\footnote{
To be more accurate the formulas in \cite{brc} refers to $n\le 1$,
however, as previously discussed, the contribution of terms
with $n\ge 2$, being higher order corrections, is extimated to be negligible.}
From now on we will refer only
to order $\alpha$ corrections,
since they are  the only relevant ones. 

In low energy 
$\nu_e(\overline{\nu}_{e})e^{-}$ scattering experiments,
an event with a
$\nu_e(\overline{\nu}_{e})e^{-}\gamma$
final state
has to be rejected because it looks like  
a multi-Compton background event ({\it i.e.}
an event due to a $\gamma$ which has more 
than
one Compton scattering inside the detector).
Therefore the experimentally relevant cross section is
given only by the sum of final states with an arbitrary number of
photons wich are not identified in the detector, {\it i.e.}, to a good degree
of accuracy, photons with an energy lower than a given threshold energy
$\omega_{min}$.
 Soft photons ($\omega<\omega_{min}$) still ensure the correct cancellation
of unphysical infrared QED divergences, however the
contribution of $\nu_e(\overline{\nu}_{e})e^{-}\gamma$ events is
lower than in the fully inclusive case.

To study the size of this effect we have built up an event
generator for the process
$$\nu_e(\overline{\nu}_{e})e^{-}
\rightarrow \nu_e(\overline{\nu}_{e})e^{-}\gamma$$

The cross section for this process has been computed in \cite{neg}.
Here we evaluate the amplitude of the process numerically
using explicit representation for spinors, gamma matrices and
the photon polarization vector. Namely we write 
the scattering amplitude, $\cal M$, as
\ba
{\cal M} & = & 2 \sqrt{2} e G_F
\bar u_e(p_f)  \Big [ \sls{\epsilon}
\frac{m_e+\sls{k}+\sls{p_f}}
{2 p_f \cdot k}\gamma_\mu  (g_L P_L + g_R P_R)  \nonumber \\
& & 
+\gamma_\mu (g_L P_L + g_R P_R)\sls{\epsilon} \frac{m_e-\sls{k}+\sls{p_i}}
{-2 p_i \cdot k} \Big  ]
 u_e(p_i) N_\mu 
\ea
where $G_F$ is the Fermi constant, $e$ the electromagnetic charge,
 $P_L$ and $P_R$ are the left and 
right-handed chiral projectors respectively, $u$ and $v$ are the conventional
Dirac four spinors, 
$N_\mu=\bar v (q_i)\gamma_\mu P_L v (q_f), \ \ 
\bar u (q_f)\gamma_\mu P_L u (q_i),$ for $\bar \nu_e$ and $\nu_e$
respectively, $p_f$ and $p_i$ are the final and initial electron
four momenta, $q_f$ and $q_i$ are the final and initial neutrino
four momenta, $k$ the photon four momentum and $\epsilon$ the
photon polarization vector.
We have checked that our results are in agreement with those in \cite{neg}.

\subsection{Reactor antineutrinos and solar neutrinos}
The threshold energy for 'detectable' photons depends on the detector.
In the following we assume a gas detector, 
for instance a time projection chamber (TPC), 
as in the MUNU 
\cite{Mun97}
experiment and in the HELLAZ \cite{seg} and SuperMUNU 
\cite{koa}
\cite{eri}
projects for the 
spectroscopy of low energy neutrinos from the Sun. In a TPC both the energy 
and the topology of the event are reconstructed.
Therefore, the bremsstrahlung photon is seen 
only if its interaction point is well separated in space from the 
track of the electron recoiling  after the neutrino scattering 
and if it gives rise to a signal above the electronic noise.
Since the attenuation length of a 10 $keV$ photon in a gas
at atmospheric pressure is always 
longer than 10 $cm$ (for instance in $CF_{4}$ is 
around 50 $cm$) then a photon threshold above 10 $keV$ would be
dictated only by the level of 
the electronic noise of the given detector.

In the following we will discuss the impact of vertex bremsstrahlung 
corrections
on various integrated as well as differential quantities. To give a 
better feeling of the size of the effect we always show cross sections
averaged over the reactor (solar) antineutrino (neutrino) spectrum normalized
to unity, {\it i.e.:} 
\be
\frac {d \sigma} {d \lambda} = \Big [ \int {\mathrm d} E_\nu \ 
 {\Phi_\nu}(  E_\nu ) \Big]^{-1}
\int {\mathrm d} E_\nu \ 
{\Phi_\nu}(  E_\nu )\frac {d \sigma} {d \lambda}(  E_\nu )
\ee
where $\lambda$ denotes a generic observable (electron energy, 
scattering angle),
$E_\nu$ is the antineutrino (neutrino) energy, ${\Phi_\nu}(  E_\nu )$
is the differential antineutrino (neutrino) spectrum and
$\sigma(  E_\nu )$ denotes the cross section for the
considered process as a function
of the incident  antineutrino (neutrino) energy.

To be as general as possible,
in fig.~\negm\ we give the cross section $\sigma_\gamma$ for the
$\overline{\nu}_{e}e^{-} 
\rightarrow \overline{\nu}_{e}e^{-}\gamma$
process, weighted over the reactor antineutrino spectrum and
plotted as a function
of the photon energy threshold for three different
values of the electron kinetic energy threshold $T_0$.
A threshold energy $\omega_{min}$ of $10$ $(50)$ $keV$ for
the bremsstrahlung photon implies a reduction of 1.2\% (0.6\%) of
the cross section with respect to the inclusive one at $T_0$=0.5 $MeV$
(this percentage becomes 0.9\% and 0.5\% for $T_0$=0.1 $MeV$).
Such a reduction is a significant fraction of the error expected
in a reactor experiment and it has to be corrected for.

We observe that the 
$\overline{\nu}_{e}e^{-} 
\rightarrow \overline{\nu}_{e}e^{-}\gamma$
cross section increases with the electron energy. As a matter of fact it is 
well known that the radiation must vanish with the electron kinetic energy 
and therefore
the bulk
of the correction is for the more energetic electrons.
This is clearly seen in fig.~\negdiff\ where the differential
 cross section for the $\overline\nu_e e^- \gamma$
final state, weighted over the reactor spectrum and plotted as a function 
of the electron kinetic energy and of its scattering angle,
 is compared with the leading order one ($\overline\nu_e e^- $
final state). 
The 
cross section ratio is shown for four different photon energy 
thresholds: 1, 10, 50 and 100 $keV$. 
The 1 $keV$ threshold is taken into account just to show the trend of the 
effect (a 1 $keV$ photon converts too close to the electron track to 
be distinguished from it). 
We see that, for
photon energies above 10 $keV$, the size of the differential cross 
section ratio 
is smaller than 3$\%$ for any electron energy, even if a 100$\%$ 
photon detection efficiency is assumed.

It is interesting to notice that the
effect of this correction on the electron energy spectrum
could cancel\footnote{We recall once more
that bremsstrahlung correction as given in fig.~\negdiff\
have to be {\it subtracted} from the inclusive 
cross section given in \cite{brc}.}
 the one induced by the neutrino magnetic moment because,
whereas radiative corrections have the
overall effect to lower the cross section, the contribution
of the neutrino magnetic moment would be additive.

We now come to solar neutrinos.
The size of the inclusive one loop radiative corrections has been calculated 
in \cite{brc} and found to be less than about 2$\%$ for the $pp$ and 
$^{7}Be$ neutrinos. Here we discuss the contribution of the
events with a vertex bremsstrahlung photon above a certain energy 
in the final state.
Fig.~\ppdiff\ shows the ratio between the differential
 cross sections for the $\nu_e e^- \gamma$
and the 
$\nu_e e^- $ scattering (both
weighted over the $pp$ neutrino spectrum)
plotted as a function 
of the electron kinetic energy for different values of the photon energy 
threshold. The same is shown in fig.~\bediff\ 
for the $^{7}Be$ neutrinos (862 $keV$ line). 
We see that the contribution of the radiation
process is smaller than 0.1\% and 0.7\% for the $pp$ and $^{7}Be$
neutrinos respectively at any electron recoil energy for
$\omega_{min}$=10 $keV$. 
As a consequence, the number of 
$\nu_e e^- \gamma$
events 
with a detectable photon in the final state 
is a negligible fraction of all the 
$\nu_e e^- $ interactions and their rejection is completely harmless in a
low energy solar neutrino experiment. As a matter of fact it amounts to
0.09\% and 0.45\% ($T_0$=0.1 $MeV$ and $\omega_{min}$=10 $keV$)
for the $pp$ and $^{7}Be$ neutrinos respectively.

\subsection{Eikonal approximation}
The energy of the photons we are dealing with is substantially
lower than the kinetic energy of the detected electrons. Usually,
in this
limit, the eikonal approximation turns out to be a good approximation.
In this approximation the amplitude for the radiative process is
written as the product of the leading amplitude times the 
eikonal current. 
For the process of interest to us
this implies
\begin{equation}
\frac {d \sigma_\gamma}{d T_e}
 = \frac {d \sigma_0}{d T_e} e^2 \frac
{d^3 k}{2\omega(2\pi)^3}
\left | \frac{ p_f}{k p_f}-\frac{ p_i}{k p_i}\right |^2
\end{equation}
where $\sigma_\gamma$ denotes the bremsstrahlung cross section,
$\sigma_0$ the leading order one, $T_e$ is the electron kinetic energy,
$k$ is the photon four momentum and $\epsilon$ the photon polarization.

In fig.~\eic\ we show the comparison of the results obtained in the eikonal
approximation versus the exact ones for the reactor antineutrinos. 
It is clearly seen
that both for total rates as well as for differential
quantities the agreement is at the few per mille level, 
{\it i.e.}~sufficiently accurate for our pourposes.

Using the eikonal approximation we are in the position
to provide a very simple formula for radiative corrections
due to real bremsstrahlung:
\begin{equation}
\frac {d \sigma_\gamma}{d T_e} =  \frac {d \sigma_0}{d T_e} \frac{\alpha}{\pi}
\left ( 
\frac{E_f}{p_f}\log\frac{E_f+p_f}{E_f-p_f} - 2 
\right ) \log\frac{\omega_{max}}
{\omega_{min}}
\label{eic}
\end{equation} 
where $\alpha$ is the fine structure constant, $E_f$ and $p_f$ are the energy
and momentum of the recoil electron, $\omega_{max}$
is the maximal photon energy  ({\it i.e.}~the initial neutrino energy)
 and $ \omega_{min}$ is the threshold energy for detected photons.

The measurable cross section is hence given by 
the formulas as in {\it e.g.~}\cite{brc} minus the contribution
of real bremsstrahlung (\ref{eic}).

\section{Conclusions}
We have discussed  
the relevance of the rejection of the 
$\overline{\nu}_e({\nu}_{e})e^{-} 
\rightarrow \overline{\nu}_e({\nu}_{e})e^{-} \gamma$
events in experiments designed 
to detect with a TPC the low energy 
$\overline{\nu}_{e}$ and $\nu_{e}$
from a nuclear reactor and from the Sun.
As a matter of fact such events have 
to be disregarded 
because they look like  
multi-Compton background events inside the detector.
We have found that their rejection has a sizeable effect in a reactor 
experiment where up to 1.2\% of the antineutrino events with
electron energy above 500 $keV$ have a photon of more than 
10 $keV$ energy. On the other hand their contribution is negligible for 
the $pp$ and $^{7}Be$ solar neutrinos (0.09\% and 0.45\% at 
100 $keV$ electron
threshold). Finally we checked that the eikonal approximation is a good
approximation (at the per mille level)
of the differential exclusive cross section. 

\vskip 10pt

\noindent {\bf Acknowledgements} It is a pleasure to thank L.~Caneschi,
G.~Fiorentini and A.~Tadsen for useful discussions.
The work of M. Moretti is funded by a Marie Curie fellowship (TMR-ERBFMBICT
971934).

\newpage
\noindent\hskip -7pt
{\Large {\bf Figure Captions}}

\vskip 20pt
\begin{list}{P}{\labelwidth=100pt} 

\item [Fig.~\rcorr] Fractional  contribution $R_{\theta,T}$
of one loop electroweak corrections to $\bar\nu_e e^-$ 
cross section
as a function of the scattering angle $\theta$ ($a$)
and the $e^-$ recoil energy $T$ ($b$).
 $R_{\theta,T}$
is defined as $R_\lambda=(<d\sigma_1 / d\lambda > - 
<d\sigma_0 / d\lambda > )/ <d\sigma_0 / d\lambda >  $, $\lambda=\theta,T$,
where
$<\sigma_{0,1}>$ are the Born and one loop 
 $\bar\nu_e e^-$ scattering cross sections averaged
over the reactor antineutrino spectrum.

\item [Fig.~\negm] Cross section $\sigma_{\gamma}$ for the 
process $\bar \nu_e e^- \rightarrow \bar \nu_e e^- \gamma$ as a function
of the photon threshold detection energy $\omega_{min}$ for three different
electron threshold detection energies: 0.1 $MeV$ (dashed line),
0.3 $MeV$ (dot-dashed line) and  0.5 $MeV$ (continuos line).
The Born level cross sections, $\sigma_0$,
for  the process $\bar \nu_e e^- \rightarrow \bar \nu_e e^-$,
are $52$, $43$ and $35$ respectively.
Units of $10^{-46}cm^{2}$ are used.

\item [Fig.\negdiff] Ratios
 among the differential cross sections, weighted over the reactor 
antineutrino spectrum,
for $ {\bar \nu_e e^- \gamma}$ and ${\bar \nu_e e^- }$ final 
states. 
Fig.\negdiff $a$, $R_{T\gamma}=(d \sigma_\gamma/d T)/(d\sigma_0/d T)$
as a function of $e^-$ kinetic energy  in $MeV$.
Fig.\negdiff $b$, $R_{\theta\gamma}=(d\sigma_\gamma/d \theta)/
(d\sigma_0/d \theta)$ as a function of the electron scattering angle.
The 
ratios are shown for four different photon energy 
thresholds: 1, 10, 50 and 100 
$keV$ 
(dotted, continuos, dot-dashed and dashed
line respectively). 

\item [Fig.\ppdiff] Ratio among
the vertex bremsstrahlung contribution and the tree-level one,
both weighted over the solar spectrum,
for the $pp$ neutrinos at $\omega_{min}=1,\ 10,\ 50, \ 100$ $keV$ 
(dotted, continuos, dot-dashed and dashed
line respectively). 

\item [Fig.\bediff] Ratio among
the vertex bremsstrahlung contribution and the tree-level one
for the $^{7}Be$ neutrinos at $\omega_{min}=1,\ 10,\ 50, \ 100$ $keV$ 
(dotted, continuos, dot-dashed and dashed
line respectively). 

\item [Fig.\eic] Ratio $R_{eik}=\sigma_{ex}/\sigma_{eik}$ 
among the order $\alpha$ 
cross sections with real bremsstrahlung treated exactly, $\sigma_{ex}$,
and in the eikonal approximation, $\sigma_{eik}$.  $R_{eik}$ is plotted
as a function of the electron kinetic energy ($MeV$). The threshold energies
$\omega_{min}$ for the bremsstrahlung photon are:
1 $keV$ (continuos line), 50 $keV$ (dot-dashed line) and 100 $keV$ 
(dashed line).
Cross sections are averaged over the reactor antineutrino spectrum.

\end{list}
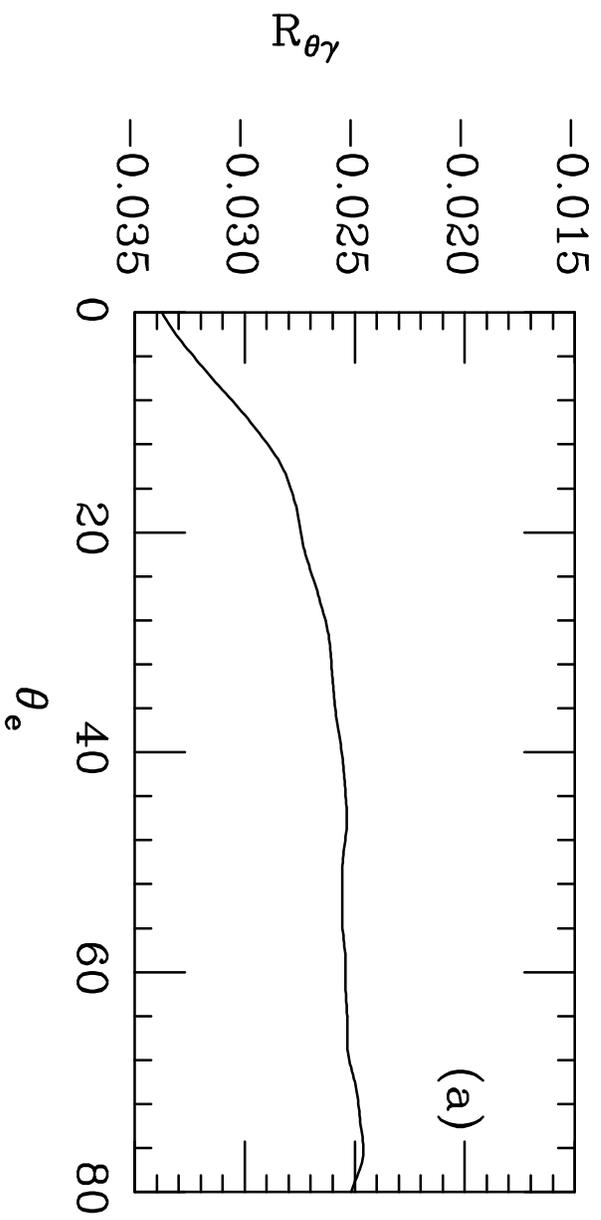
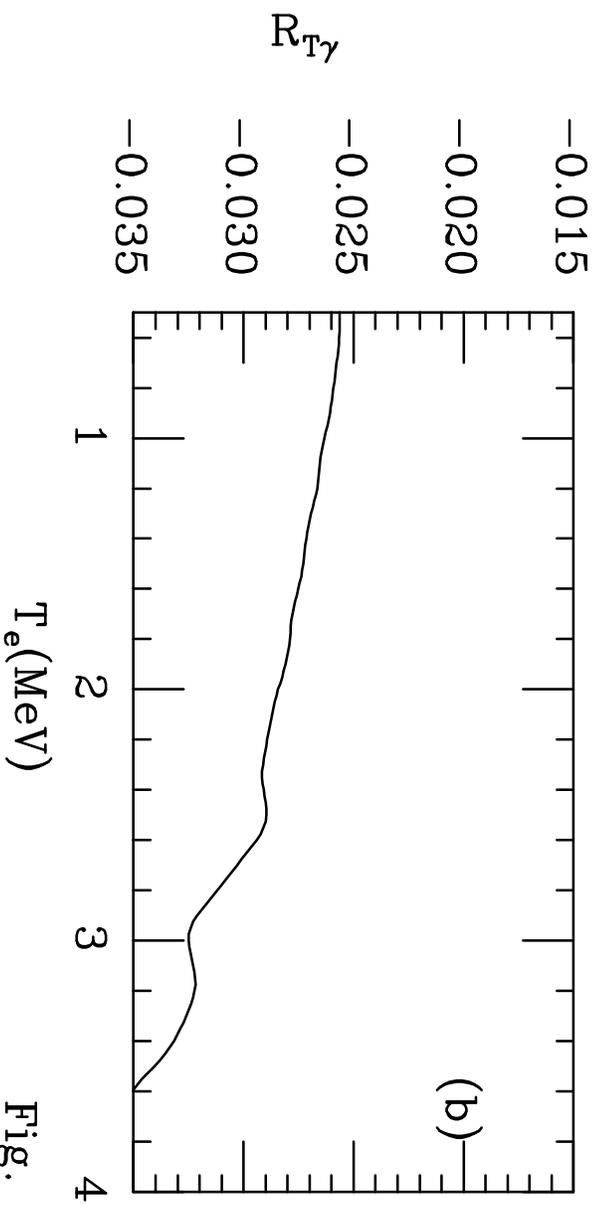
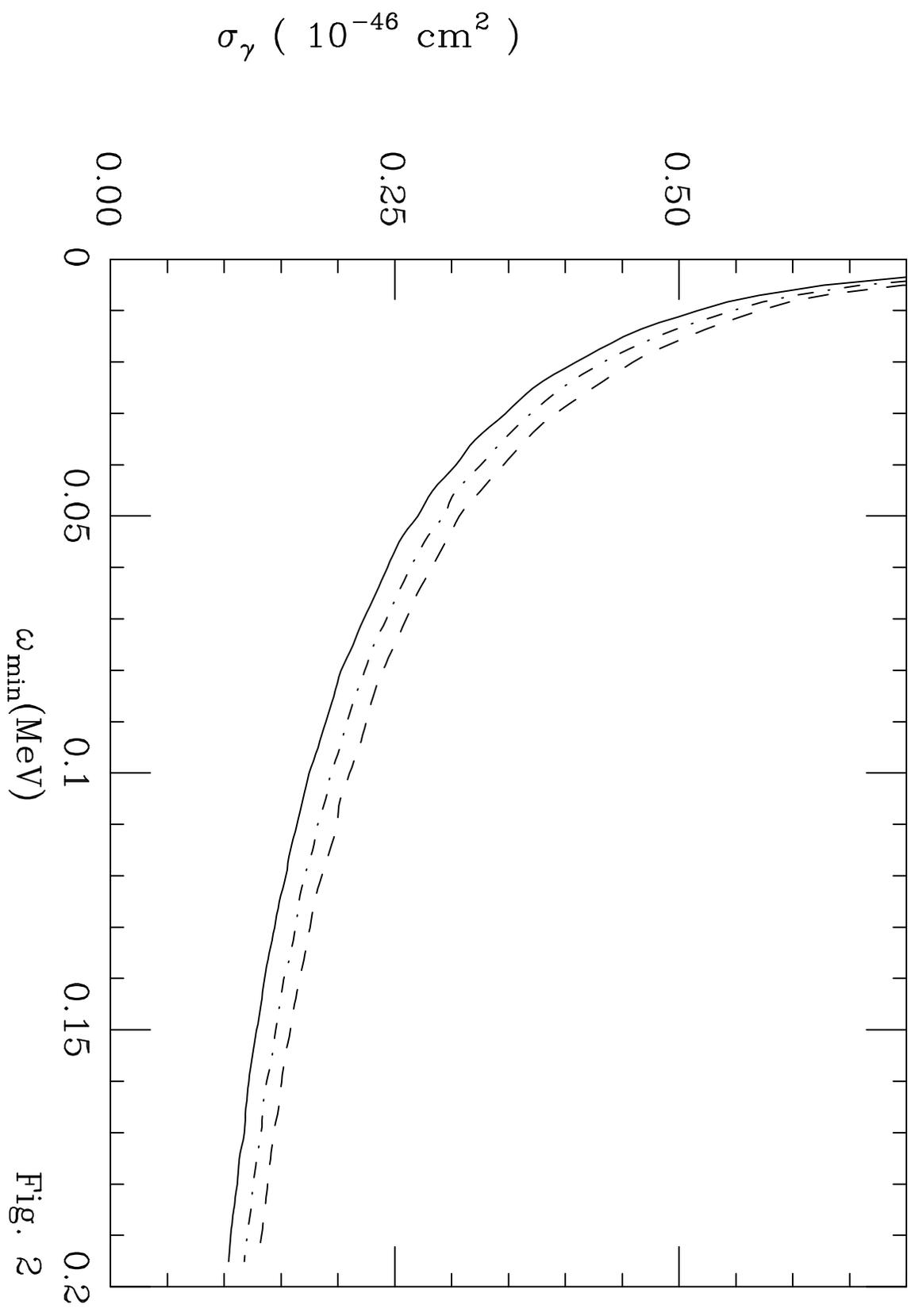
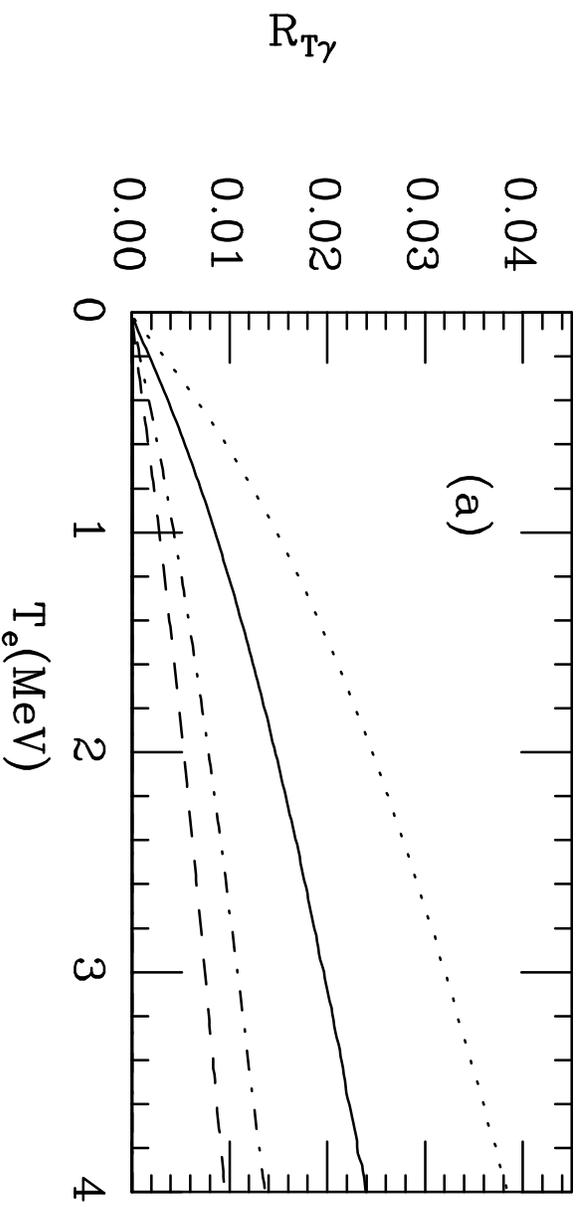
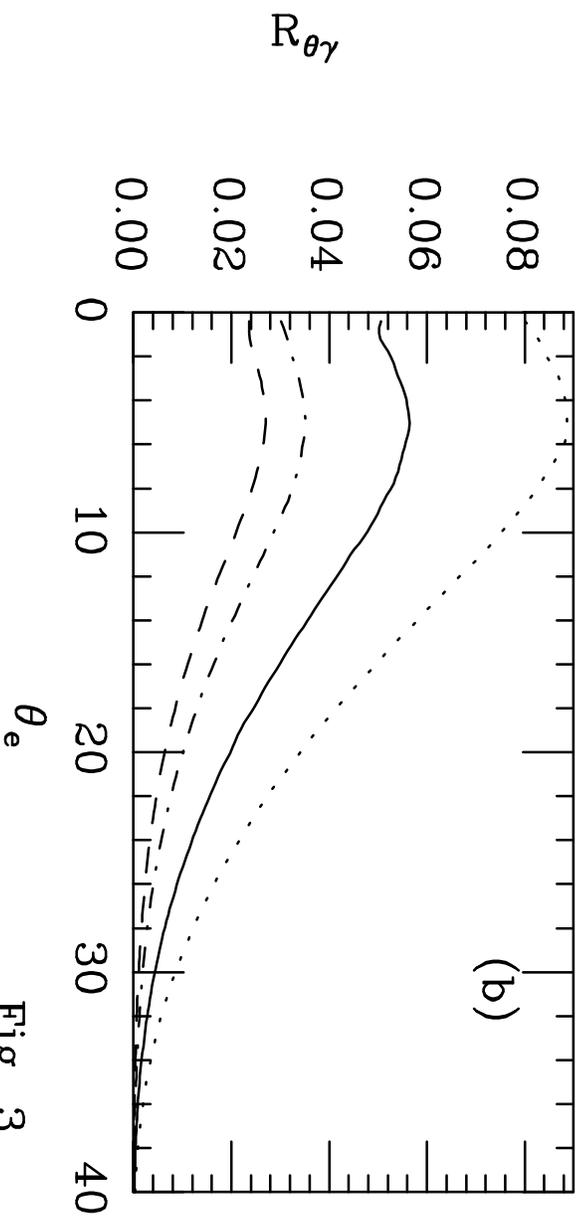
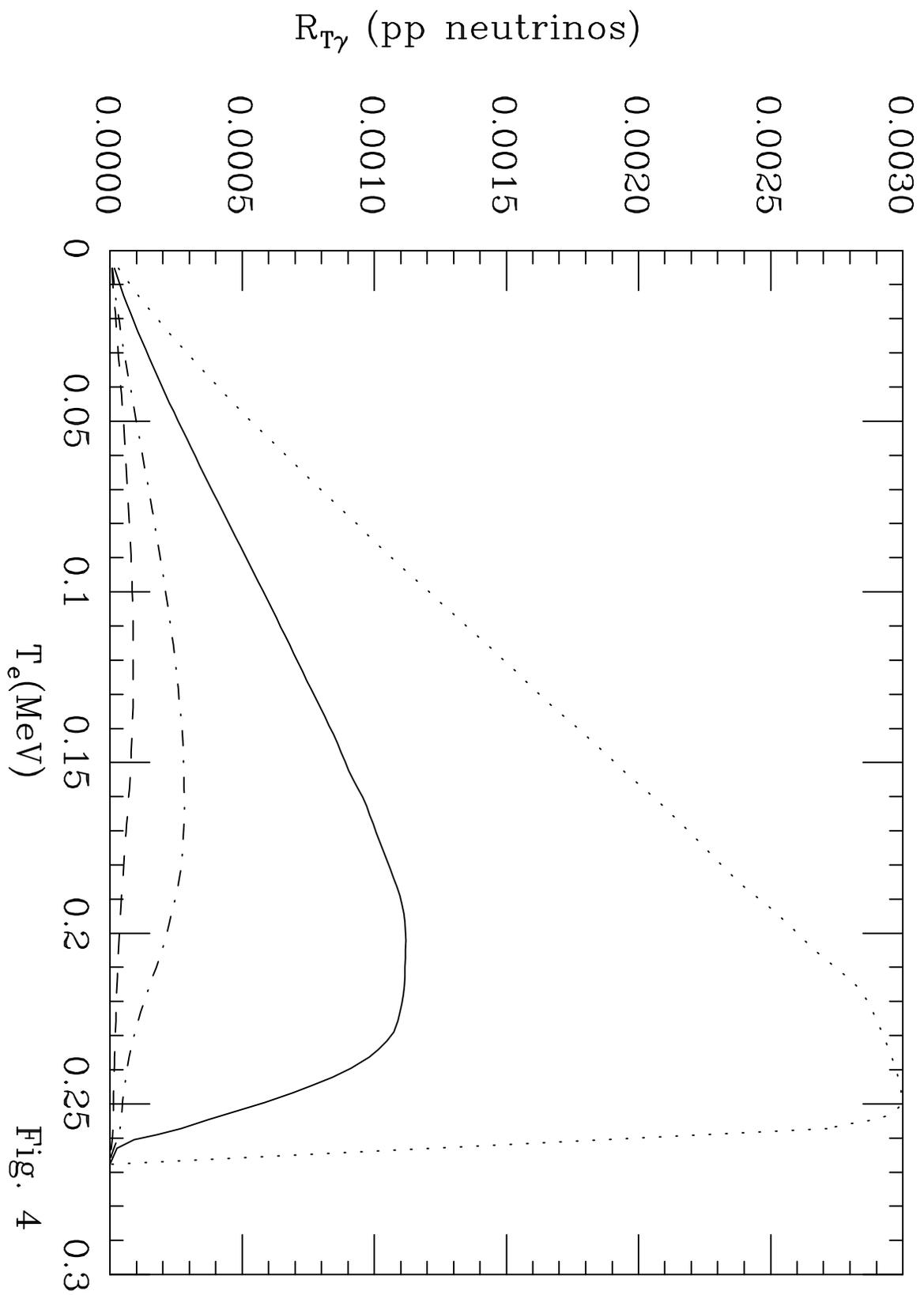
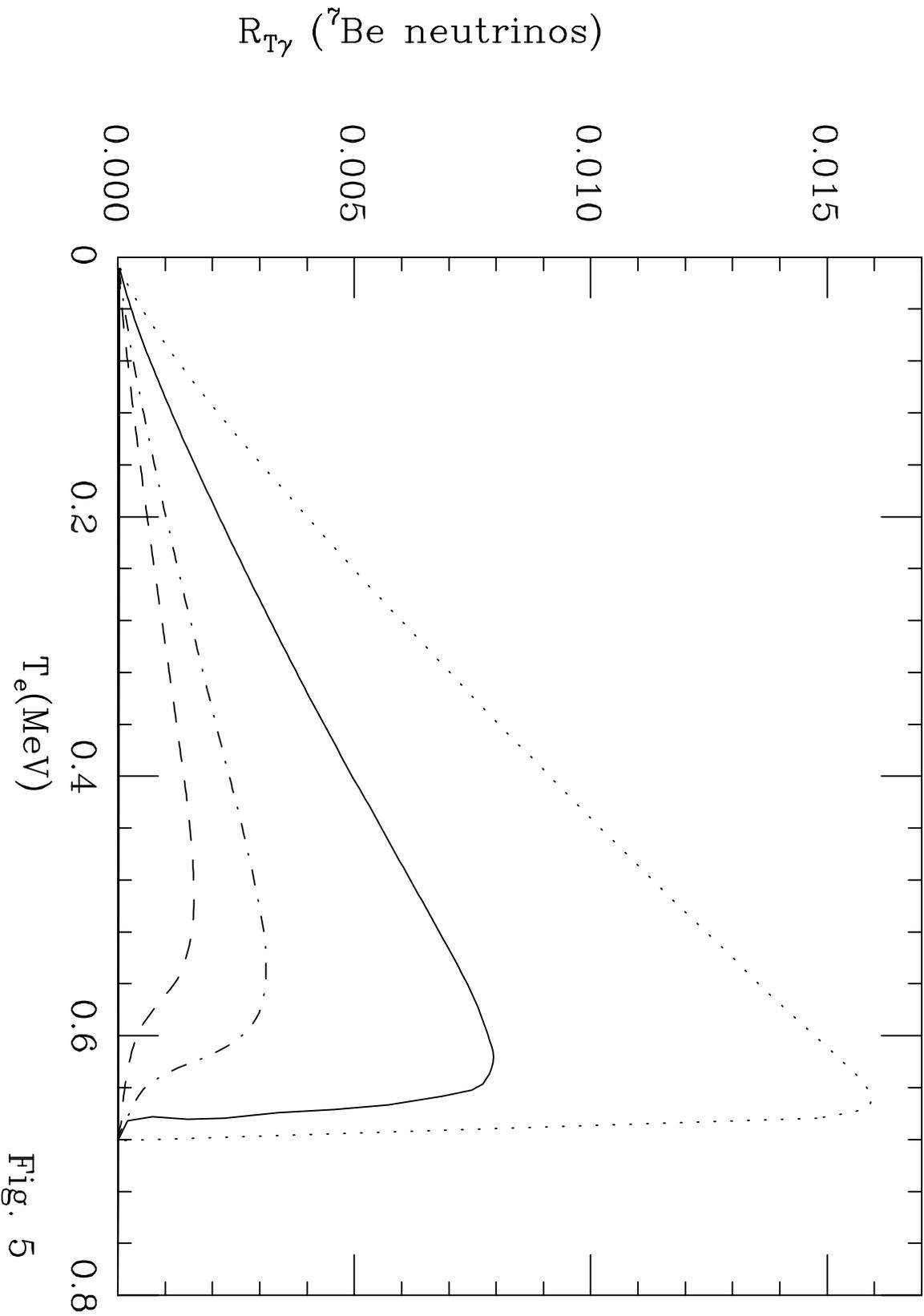
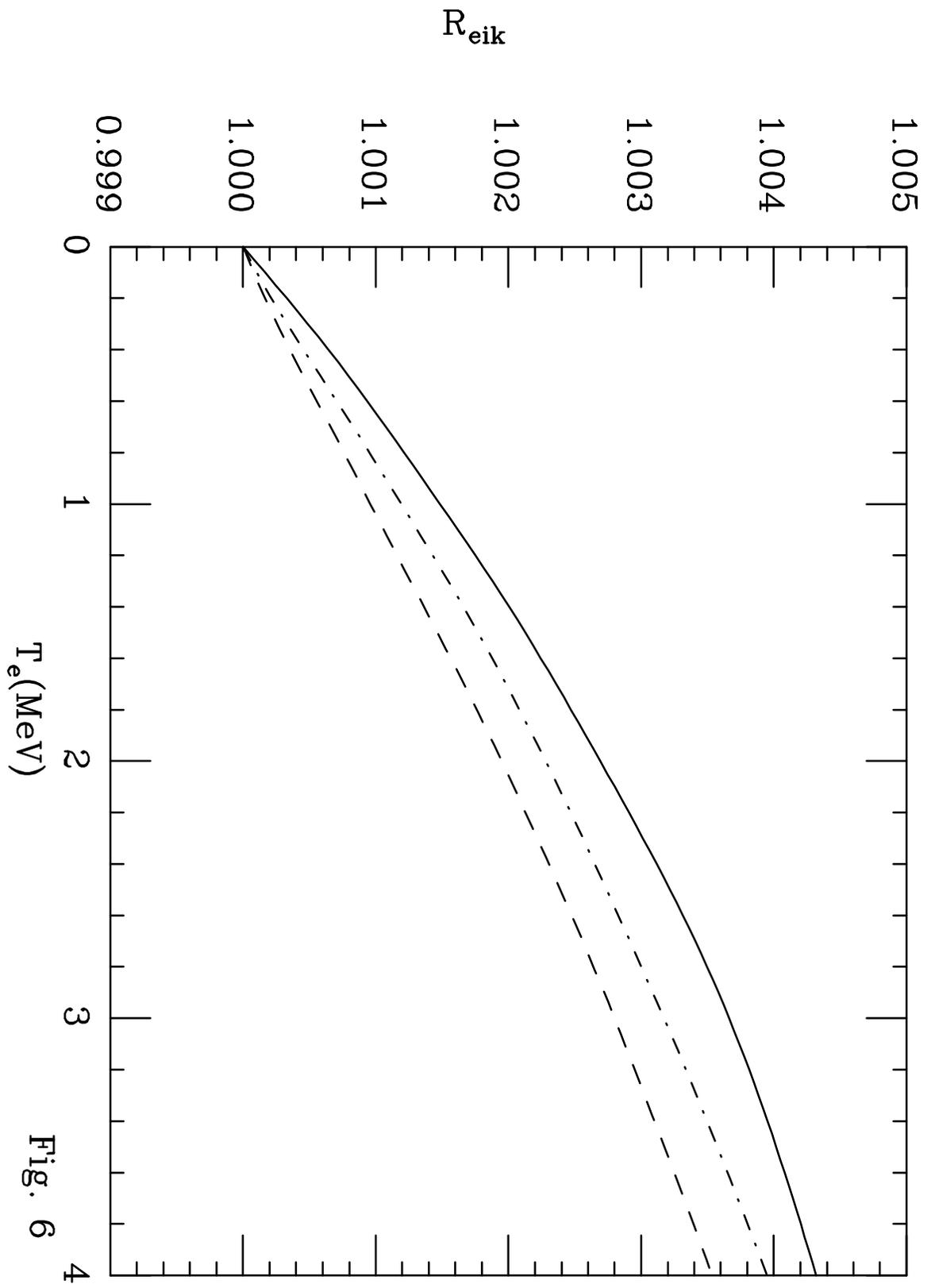
\end{document}